# A Robust Adversary Detection-Deactivation Method for Metaverse-oriented Collaborative Deep Learning

Pengfei Li, Zhibo Zhang, Ameena S. Al-Sumaiti, *Senior Member, IEEE*, Naoufel Werghi, *Senior Member, IEEE*, and Chan Yeob Yeun, *Senior Member, IEEE*

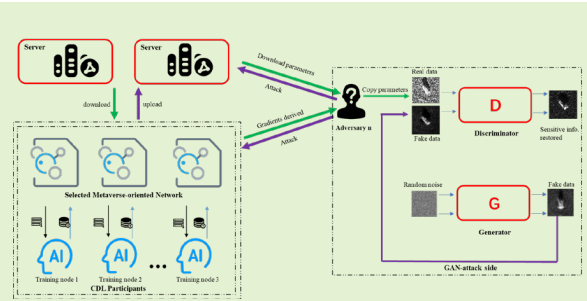

***Abstract*—Metaverse is trending to create a digital circumstance that can transfer the real world to an online platform supported by large quantities of real-time interactions. Pre-trained Artificial Intelligence (AI) models are demonstrating their increasing capability in aiding the metaverse to achieve an excellent response with negligible delay, and nowadays, many large models are collaboratively trained by various participants in a manner named collaborative deep learning (CDL). However, several security weaknesses can threaten the safety of the CDL training process, which might result in fatal attacks to either the pre-trained large model or the local sensitive data sets possessed by an individual entity. In CDL, malicious participants can hide within the major innocent and silently uploads deceptive parameters to degenerate the model performance, or they can abuse the downloaded parameters to construct a Generative Adversarial Network (GAN) to acquire the private information of others illegally. To compensate for these vulnerabilities, this paper proposes an adversary detection-deactivation method, which can limit and isolate the access of potential malicious participants, quarantine and disable the GAN-attack or harmful backpropagation of received threatening gradients. A detailed protection analysis has been conducted on a Multiview CDL case, and results show that the protocol can effectively prevent harmful access by heuristic manner analysis and can protect the existing model by swiftly checking received gradients using only one low-cost branch with an embedded firewall.***

***Index Terms*— Metaverse, collaborative deep learning (CDL), GAN-Attack, adversary detection, adversary deactivation, privacy protection.**

## I. Introduction

**K**NOWN as an innovative concept first proposed in the book (Snow Crash) by Neal Stephenson [1], the metaverse has been an idea that roots inside the public mind and become a good wish throughout these decades. Back to its advent, the metaverse was initially considered too fancy to be real and had lived in science fiction as an 'illusion concept'. However, the fast growth of Artificial Intelligence (AI) and those powerful Deep Learning (DL) methods have made constructing such a world possible as these techniques can process huge amounts of data even under a time restriction. This impressive attribute matches the real requirement of the metaverse which desires fast computation ability to map the real world to a digital environment (such as 3D shopping in a virtual mall), so as to achieve the goal of being interactive, cooperative, and immersive. Therefore, the realization of the metaverse would require building a deep effective model for data processing, and this work may require various teams or participants to cooperate together to make the model be trained on more data and, thus, achieve better results with enhanced generalization ability and accuracy. This approach is known as Collaborative Deep Learning (CDL) with a parameters aggregation process [2]–[5]. Another advantage brought by CDL is the isolated nature of training data, as CDL only requires different groups to upload gradients derived from their own data rather than the data itself. This nature can realize enhanced security in private data protection. Traditionally in CDL, people are attracted by the effectiveness of this cooperation manner in training a better model, while seldom realizing that the CDL is actually exposed to some secure menaces [6]–[8]. However, as pointed out by Aono *et al.* [9], CDL can be compromised by the server even when only gradients are uploaded, and this breach of gradient usage can make privacy at risk. To strengthen the protection, they adopted additional homomorphic encryption (HE) to protect gradients to be uploaded. Also, Geyer *et al.* proposed a differential

Pengfei Li (e-mail: pengfei.leen@outlook.com), and Zhibo Zhang are with the Department of Electrical Engineering and Computer Science, Khalifa University, Abu Dhabi, UAE.

Ameena S. Al-Sumaiti is with Advanced Power and Energy Center, Electrical Engineering and Computer Science Department, Khalifa University, Abu Dhabi, UAE

Naoufel Werghi is with Department of Electrical and Computer Engineering, Khalifa University, Abu Dhabi, UAE

Chan Yeob Yeun is with Center for Cyber-Physical Systems, Khalifa University, Abu Dhabi, United Arab Emirates (e-mail: chan.yeun@ku.ac.ae).

privacy (DP)-based method to constraint illegal data usage from the server.

In 2017, Hitaj, Ateniese *et al.* [10] found that the generative adversarial network (GAN)-based technology even makes the participant side able to launch an attack that can retain private data from downloaded gradients, and this GAN-type attack manner, due to its capability, attracts many investigations in recent years [11]–[14]. Under this new threat, Chen *et al.* [15] proposed a Trusted Third Party-based protocol to separate the innocent participant and the server so that none of these two sides can restore sensitive input from gradients. Also Yan *et al.* [16] used a buried point, in conjunction with a further protection action, to defeat the participant-side attack, and they demonstrated the protocol effectiveness in a case when GAN-attack is disabled by adjusting the learning rate.

This paper uses a metaverse-oriented multiview (MV) model [17], [18] as an investigation case and targets providing more security to applications by adopting a new protocol that can protect the CDL process. The detailed application scenario can be various, as one example, it can be providing privacy protection to metaverse 3D-immersive shopping in a virtual mall. To realize such a concept, numerous sensors are required to scan and rebuild details of a shopping mall's layout and commodities such as clothes. More importantly, there will be a pre-trained model (just like the model in CDL server) that can generate a virtual dressed human using the technique of few-shot learning [19], [20] to make a customer exactly know what he/she will be like once get dressed with selected clothes. This process requires customer (just like an innocent participant in CDL) to upload some private data such as bust measurement. Therefore, it is of practical motivation to develop an algorithm to strengthen the privacy protection in metaverse-oriented CDL.

This paper divides threats to metaverse-oriented CDL into two main categories: 1) The malicious behaviors toward the model itself, and 2) The malicious intentions toward the sensitive information from innocent participants. In detail, the first point refers to the fact that adversaries may try to degrade or even destroy the model by uploading some useless or wrong gradients. While the second point: attacks on innocent participants, can be launched either from a bad server or a bad participant. For the second category, since there are already many mature investigations on quarantining a harmful server [8], [9], [21], this paper primarily focuses on defeating dangerous participants who apply GAN-based attack technology. The contributions of the paper can be summarized as follows:

1) This paper classifies the major five types of threats in CDL and tries to provide the metaverse-oriented CDL system with comprehensive protection by designing a new protocol.

2) The paper identifies a GAN-attack by heuristic scanning and later limits or denies the request from the suspicious CDL node.

3) The paper applies an embedded adversary detection node as a signature that can detect malicious gradients uploaded by dangerous participants.

4) The paper applies a firewall that can quarantine the malicious gradients and therefore, deactivates attacks that aim at existing half-trained CDL models in the server.

5) The paper conducts a comprehensive protection analysis with results being displayed via a useful expandable Multiview (MV)-CNN.

The reminder of this paper is organized as follows: in section II, a detailed description of CDL and brief mathematical deductions of the GAN-attack working principals with gradient descent algorithm are given. These deductions are mainly used to analyze the weakness of the original CDL towards GAN-attack as well as provide corresponding protections that can disable GAN-attack by destroying its working chain. Section III thoroughly describes GAN-attack in CDL and categorizes all attacks in CDL into five main categories. Section IV provides the protocol for protecting the CDL with detailed explanations, and a step-by-step protection analysis is given to prove the effectiveness of this protocol theoretically. Section V uses an MV-CNN to verify the protocol piratically and give relevant results to support the conducted precious analysis. Finally, a conclusion is provided in section VI.

## II. BACKGROUND AND PRELIMINARIES

### A. Generative Adversarial Network (GAN)

Generative Adversarial Network (GAN) [22] is a kind of technique that is capable of reproducing a specialized result from the noise by loading relevant parameters through mutual gambling between its generative and discriminative hands. Generally, this method can be useful in augmenting limited data sets (data generation) [22], [23], image translation [24], and target detection [25].

When looking into GAN, it can be realized that the structure can be divided into two modules: a discriminative model and a generative model. The generator generates new data samples from randomness while trying to imitate the distribution of real samples. While, the discriminator, which works as a binary classifier, can distinguish the produced samples from genuine samples [26]. The goal of GAN optimization, which is a minimax game process, is to reach Nash equilibrium [27] when the generator is taken to accurately apprehend the distribution of actual samples. The final goal of GAN is to make a fake image as true as possible and such that it cannot be discriminated by the discriminator (i.e. the confidence value for both real and fake images are 0.5).

Fig. 1 depicts the GAN's computation process and organizational structure. In this figure, the original data x and noise z are random variables that serve as differentiable functions' inputs for D (Discriminator) and G (Generator) respectively. The sample created by the generator (adhering to the probability distribution of actual data (P_data)) is represented by G(z) and the domain of the original data x is the same as the range of G(z). For the discriminator, it should classify the input as true if it originates from real data x. On the other hand, if the input originates from G(z), the discriminator should label it as false. The goal of the discriminator is to accomplish accurate data source categorization, whereas, the generator's aim is to guarantee consistency between the performance of generated data G(z) on discriminator D(G(x)) and the performance of actual data x on the discriminator

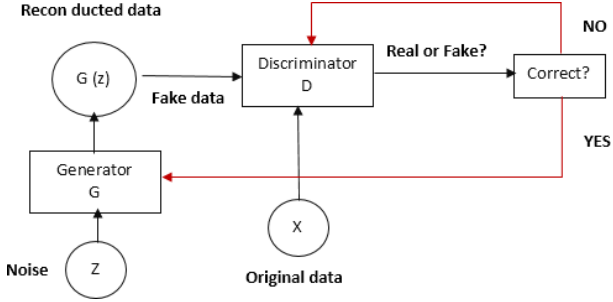

Fig. 1: The Schematic of a Typical GAN model.

D(x). In addition, D is trained to increase the likelihood that the generated and real samples will be appropriately assigned, allowing the training of the generator simultaneously by diminishing the term log(1-D(G(z))). In other words, the value function V(G, D) is the object of a minimax game between the generator and discriminator, which is the goal of GAN optimization as mentioned before.

Above adversary process can be briefly described by Eq.(1), which shows how the network tries to minimize the generator loss and maximize the discriminator loss.

$$\min_G \max_D V(D, G) = E_{x \sim P_{data(x)}}[\log D(x)] + E_{z \sim P_{z(z)}}[\log(1 - D(x))] \quad (1)$$

Where the probability distribution of the potential variables and observed data are represented by $P_z(z)$ and $P_{data}(x)$. The first component in the equation corresponds to the entropy of x when it is loaded by the discriminator, while the latter component pertains to the entropy of z when it undergoes the processing of the generator [2] [15].

*B. Collaborative Deep Learning*

The advent of deep learning (DL) brings great changes to people's daily lives; however, the growing size of DL-related networks, as well as the parameters that need to be trained, makes the traditional centralized training mode more difficult [5] [28]–[30]. This fact indicates that involving more participants (or contributors) are the real need as collaboration makes the work better, and also, more participants indicate more specialized or build-up data sets, which are always desired by the AI specialist as the generalization ability of the network can be improved. Moreover, a transition from centralized training to semi-centralized to decentralized is gradually clearer. For example, before 2014, most neural networks, such as LeNet [31], AlexNet [32], were proposed and trained locally. Later, the well-known Google VGG-16 network [33] was proposed, which directly led to a fashion of building pre-trained models. The VGG-16 itself has a pre-trained version based on the ImageNet [34] data set, which was continuously built up through collecting multiple images from different sources. This stage can be seen as semi-centralized network training as the data sets vary diversely rather than being private but the training process still remain a secret. Finally, with the increasing real need for collaborative work as well as the increasing concern in data privacy, the CDL becomes much more common [7] [16]. This new collaborative format stresses both the requirements of 'more contributors' and 'better data privacy protection'. Although sometimes the CDL might require more administrative communication than the centralized training, the decentralization (or localization) nature addresses the valued privacy concerns by isolating participants and also, by avoiding directly uploading sensitive local information [28], [29].

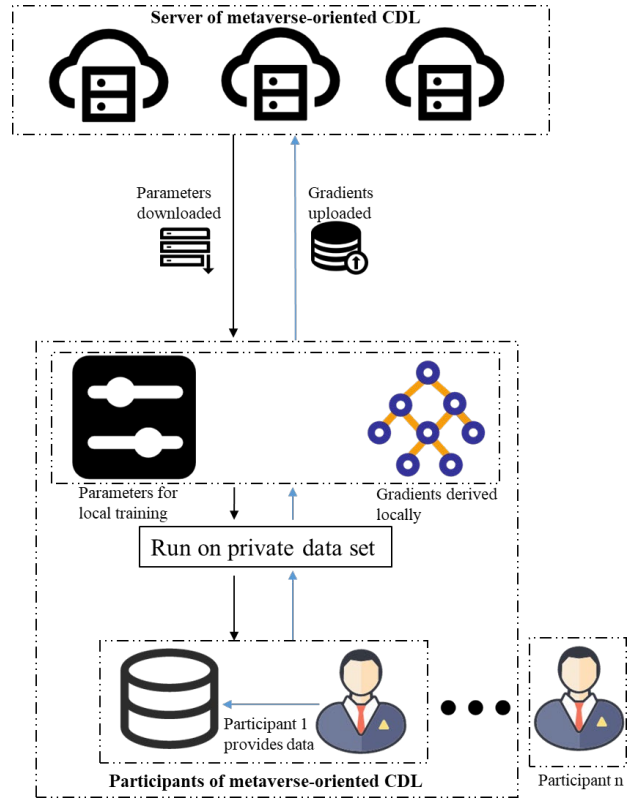

Fig. 2: Detailed process of a typical CDL.

In collaborative deep learning (CDL), a man can train the model in his localized data while only requiring to upload the derived gradients to the global parameters to further update the model weights. Fig. 2 illustrates a typical format of CDL by listing one participant as an example. Through this figure, we can firstly find two parts: a global server from where up-to-date parameters were downloaded and loaded into the model as a good start of localized training and the participants who continue to train the model with his/her local data set to get different gradients for network weights that are updated in a later back propagation (BP) process. Later, the valued gradients can be uploaded to help refresh the global parameters without the necessity of providing the participant's private data.

*C. The Cross-Entropy Loss and Gradient Descent for Minimizing the Error*

For refreshing the weights and biases and looking for optimized values of network parameters, the Stochastic Gradient Descend (SGD) optimizer (or its further development such as the Momentum, and Adam optimizers) is applied to minimize the error between predicted results and real results. It is noticeable that different tasks rely on different kinds of loss functions, for instance, the multi cross-entropy loss function is applied in multi-categories classification tasks, the mean square error (MSE) is usually for predictions and detection task relies more on a loss function associated with a bounding box, and segmentation task may need a loss function that assigns more weights to the boundaries of interested area. Here, we use the commonly-used cross-entropy loss as an example, where $CE(w^k, b^k)$ denotes the multi-categorial cross-entropy loss associated with the last layer where the backpropagation starts from. The refresh process of weight and biases can be denoted as w and b as shown by Eqs.(2) and (3), respectively.

$$w^{k+1} = w^k - \mu \times \frac{\partial CE(w^k, b^k)}{\partial w^k} \quad (2)$$

$$b^{k+1} = b^k - \mu \times \frac{\partial CE(w^k, b^k)}{\partial b^k} \quad (3)$$

Note that $\mu$ means the learning step, $w^k$ and $b^k$ represent the old weight and bias value of the last layer, and $w^{k+1}$ and $b^{k+1}$ are the refreshed weight and bias value, respectively.

For the gradient descent process, let us take 'w' as an example. Since w is an old known parameter, and $\mu$ is a super-parameter defined by us, the refresh of a parameter like 'w' is to solve the partial derivative of the error to the parameters (w and b). And in the last layer of the computation process where backpropagation starts and refreshes the weight, it can be found that the only unknown partial derivative can be expanded by the chain rule, here the weight is listed as an example:

$$\frac{\partial CE(w^k, b^k)}{\partial w^k} = \frac{\partial CE(w^k, b^k)}{\partial out} \times \frac{\partial out}{\partial in} \times \frac{\partial in}{\partial w^k} \quad (4)$$

where $\frac{\partial CE(w^k,b^k)}{\partial out}$ describes the back to the cross-entropy loss function, $\frac{\partial out}{\partial in}$ relates to the activation function. The sigmoid function ($out = \frac{1}{1+e^{-in}}$) in this study of CDL-targeted network can be further deducted as $out \times (1 - out)$, and $\frac{\partial in}{\partial w^k}$ is from the summation in= $\sum w^k x_i + b^k$.

## III. GAN ATTACK IN METAVERSE-ORIENTED CDL

Although to some extent the CDL protects data privacy by only uploading the gradients rather than the data itself, it is actually still under the threat of GAN attack. Like what has been introduced in section II, the GAN can even re-generate the data once the parameters are known by simply replaying the discriminative model to 'teach' the generative model so that a result that is very similar to the original input can be generated by the input noise. Therefore, in terms of metaverse-oriented CDL, the GAN threats can still do great harm to vulnerable innocent victims.

The current multiple threats can be classified as the following types:

1) An adversarial server which illegally collects private information

2) An adversarial participant who pretends to be a part of the project while in fact trying to download the weights from the global parameters and then re-produce the sensitive information by GAN-attack

3) An adversarial participant destroying the training process by uploading the incorrect gradients derived from a training of uncorrected/irrelevant images or deliberately assigning an image with a false label, or the gradients uploaded can lead to gradient vanish or explosion problem

4) An adversarial adding some non-existent classes to the categories and hence dampening the network performance in the backpropagation process

5) An adversary intending to upload malicious parameters to the established model

Fig. 3 illustrates one typical GAN attack aimed at the CDL, from which it can be observed that in this kind of attack, the server connects both the victim and the adversary where in CDL, victims download parameters and load them to the model, which will later be trained by the sensitive data sets (*A*, *B*) to get the gradients for upgrading the weights. When gradients come back to the server, the adversary can later download those parameters and feed them into the same network as the discriminator. Later, the generator, which is like the Siamese of the discriminator can be 'taught' by the discriminator so as to generate fake but like real data to make the adversary get an output that is as close as the real input and therefore, the sensitive information (A, B) are inexplicitly stolen. And even after the adversary successfully steals the input, they still want to attack more by uploading the bad gradients to the server.

## IV. ADVERSARY DETECTION AND DEACTIVATION

The core step of preventing a GAN attack in metaverse-oriented CDL is to quarantine potential attackers from the parameters once the threat or adversarial behaviors are detected. Moreover, there can be a 'pop-up' window (feedback) for the server to isolate the adversary by denying the sharing.

### A. System Architecture

The basic structure of our adversary detection-deactivation system is shown in Algorithm 1, where it is initially assumed that the server cannot distinguish what is a good participant and what is a malicious one, and hence allocate similar priority to all of them in the preparation stage. The access of different users is divided in accordance with the training epochs (for instance, epoch (1, h) for user1, epoch (h, m) for user 2, and so forth). During the local training stage, the server is unable to monitor or control participants. However, to ensure the protection to those innocent entities, access permissions are limited by $C_{at}$ which is a credential that restricts users' access trials. This credential-enforced method can be considered as a kind of heuristic scanning which will regard participants who frequently ask for new gradients as adversaries and therefore, temporarily block their request until the certificate being re-issued. After this scanning, all innocent participants and survived adversaries can continue to use downloaded parameters to generate gradients to be returned (Survived adversaries here refer to malicious nodes which aim at destroying the established model rather than reproducing others' input so that they can not be identified via heuristic scanning as they do act like the innocent until returning back gradients).

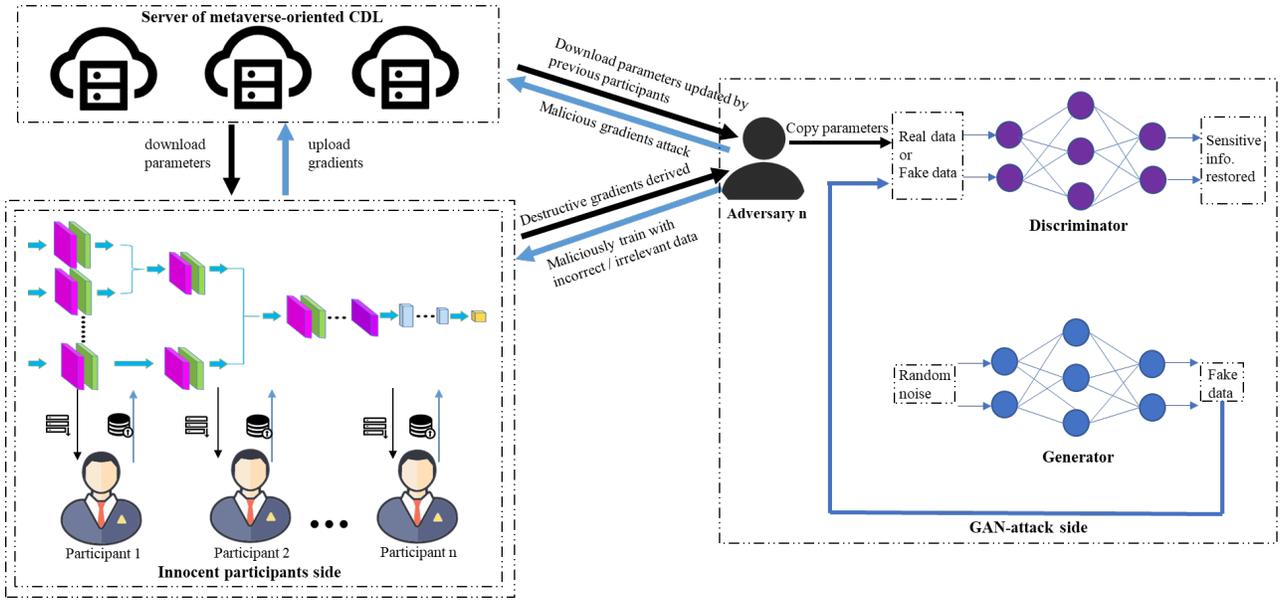

Fig. 3: A typical GAN-attack in CDL and dangerous gradients upload.

After completing the local training, 'CDL global update stage' can deliver an extra protection to the established model which is safeguarded by the check node ($N_a$) and the firewall layer ($W_f$) with weights of all '1' and biases of all '0'. Moreover, because all participants are independently assigned into different epoch slots, the malicious node can also be identified once the server observed a performance deterioration in the fast check phase (module is shown Fig.4). It is noticeable that, the fast check method works as a following-up procedure after gradients passing through firewall, and this check will only run on a model with pre-trained weights so that the derived output is the fine-tuned result from either $V_u$ or $A_u$. And the derived output can later be compared with the benchmark performance to justify the effectiveness of received gradients. In short, the fast check phase can be described as 'server activate one branch - back propagate with $V_u/A_u$ -fine tune pre-trained weights - compare derived results with benchmark results'

Finally, suspicious gradients reported by either check node or fast-check will linked with a pop-up warning, and those gradients will be further forwarded to the control center for manual review. A permanent access repudiation will be assigned if adversaries are finally identified. In case that gradients have passed through all checks, the firewall will be deactivated to make gradients accessible to the real back propagation process. More details about algorithm 1 will be went through in combination with adversary detection and anti-attack defense analysis.

**Algorithm 1: Adversary Detection and Deactivation against GAN Attack**

**Prepare Stage:**
(a) Subjects be divided into Adversaries (A) and Victims (V), existing CDL architecture in Fig. 3, V and A declare different label item sets
(b) Set a self-decaying learning rate
(c) Parameters Downloaded by V and A ($V_d$, $A_d$)
(d) Gradients Uploaded by V and A ($V_u$, $A_u$)
(e) Adversary Detection node ($N_a$), returned node ($N_a^{'}$)
(f) Access Credential with limited trials to each participant ($C_a t$)
(g) Firewall layer weights ($W_f$)

**CDL Local Training Stage (Under Server Access Control):**
1) for epochs in ranges (1, h), (h, m):
2) Check the credential validity and prevent repetitive access to parameters
3) Accept the access of innocent user V, adversary A in different epoch slots
4) V downloads $V_d$, A downloads $A_d$ from the server with embedded $N_a$
5) Load $V_d$, into local models, load $A_d$ into attack model
6) V tries to upload $V_u$ (beneficial) to the server, while A tries to upload $A_u$ (malicious)

**CDL Global Update Stage (Server Check):**
1) for epochs in ranges (1, h), (h, m), server check:
2) **Check node & Firewall Decision**
   ($N_a^{'}$)? ($W_f = drop(W_f, 1)$) : ($W_f = drop(W_f, 0)$)
3) **Extra check:** fast-check with one-branch & rate performance
4) **if deterioration observed:**
5) $W_f = drop(W_f, 1)$ # firewall on

6) **Pop-up warning:**
7) Warnings to server → restrict data transmission rate further notice specialists
8) **Adversary confirmed by specialists:** Deny A's access forever
9) **if No change in node & performance:**
10) Participant Rate → different confidence
11) Deactivate the firewall: $W_f = drop(W_f, 0)$ → bypass drop out → back propagation (BP) on

### B. Adversary Detection

From Eq. (1), we can know that a successful GAN attack is based on the continuous interaction between the discriminator and generator to reduce the loss, this continuity relies on frequently downloading and uploading parameters, which can be a feature of the malicious user. Therefore, to terminate the possible attack, we can heuristically identify the adversary, send the warning message back to the server, and block the access of illegal participants. Besides, as mentioned in section III, apart from stealing sensitive information, the adversary may also try to upload some undesired gradients to impact or even destroy the model. To deal with this kind of attack, a firewall can be set to the model according to Eqs.(2)-(4), which show that SGD relies on partial derivative and chain rule. This dependence indicates that if the gradient transmission is blocked, both model and innocent users can be protected. More specifically, block gradients means (for the adversary) other users' sensitive input cannot be reproduced, (for the model) established good parameters can be intact. In [28], the authors realize this block by adjusting learning rates because they find that neither large nor small learning rates can make the GAN model reach the optimum as the generative loss can not converge. In [15], the author adds a trusted third party to provide the anti-GAN attack service. This work investigated both two solutions and proposed the paper's own algorithm.

In this paper, instead of tuning the learning rate, a branch of 'firewall' is added, which mainly consists of a 100% dropout layer as this kind of method is more powerful and straightforward than adjusting the learning rate. According to Eq.(4), when gradients cannot be updated in the back propagation, adversarial attack can be blocked before it does harm to either the half-trained model in the server or the innocent entity. For Threat Types 3-5, these threats can be attenuated by running gradients from participants in a blank model. Or more efficiently, when running with the Multiview (multi-branch) structure, gradients can be swiftly checked from model performance by activating only one of existing branches (shown in *Fast check branch* of Fig.4). If a significant performance deterioration is observed, the malicious behaviour can therefore be defined. It is noticeable that the latter manner can outperform the first (blank model check), as with one branch, the model can be swifter in checking because of the lower computational cost and reduced running time. This conclusion will be further demonstrated in section V, part D.

### C. Anti-Attack Defense Analysis

The core methodology of this algorithm is to conduct heuristic scanning and to look for adversary-like behavior such as frequently asking for data from the server (so as to continuously improve the discriminator and teach the generator) [35]. To protect the sensitive information from innocent participants, the adversary's action of repeatedly accessing the most up-to-date parameters should be blocked. In this term, each participant is assigned a unique credential $C_{at}$, which contains the possible trails a participant can possess to download parameters. As mentioned before, one of the typical features of adversaries is that they always want to repeatedly access the server to construct a better GAN model. In this case, $C_{at}$, which can limit the access times, can realize the purpose well. This point addresses the type 2 attack in section III. It is noticeable that the type 1 attack is not evaluated in this paper as the server are much easier to be regulated in comparison with distributed participants [2].

**Algorithm 2: CDL-oriented Data Formation & MV Architecture**

**Class Cyber (Dataset):** # The MV hierarchy for data loading
  **def__init__(if** train == True):
    super(The Data, self).__init__()
    **if** train==True:
      branch 1 = data ← $1^{st}$ branch view
      branch 2 = data ← $2^{nd}$ branch view
      branch 3 = data ← $3^{rd}$ branch view
      ... ...
      branch N = data ← $N^{th}$ branch view
      Train_label = Loading (N branches ← 1-Label)
    **else:**
      branch 1 = data ← $1^{st}$ branch view
      branch 2 = data ← $2^{nd}$ branch view
      branch 3 = data ← $3^{rd}$ branch view
      ... ...
      branch N = data ← $N^{th}$ branch view
      Test_label = Loading (N branches ← 1-Label)
    **End**

  **def__len__(self):**
    **return** len ($1^{st}$ branch data)

  **def__getitem__(self, index):**
    Concatenate: $\{branch1 \times conf1, branch2 \times conf2 ...branchN \times confn\} \times$
    Build: Input ← append$\{$branch 1 to N $\}$
    Get: Test / Train label
    Dimension Match:$\{$label, input$\}$ ← MV Dimension
    **return** Tuple <input, label>

For types 3 and 4 attacks (where the adversary intends to attack the model itself by either uploading malicious gradients or adding some new categories), the Adversary Detection node ($N_a$) can also perform well. This form of embedded node is flexible, for instance, it can be an extra tensor that only concerns about the dimension of the model. Take the ten-class multi-category classification problem in this paper as an example, the interested targets are gray-scale images of 100×100. Therefore, the Adversary Detection node ($N_a$) can be in [10, 1, H, W] where 10 is the embedded dimension restriction that prevents some irrelevant categories from being added. Also, 1 in the second place restricts the type of the image to be grayscale. These restrictions can saliently increase the cost of type 4 attack. This matrix match method can be applied as a quick check method of intrusion since further inspection can be performed by running a pre-trained model in a fast-check manner to make the server knows what has been changed.

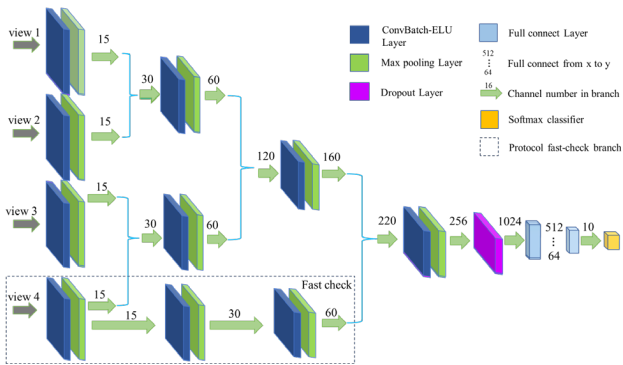

Fig. 4: The Expandable Multiview-CNN structure in CDL.

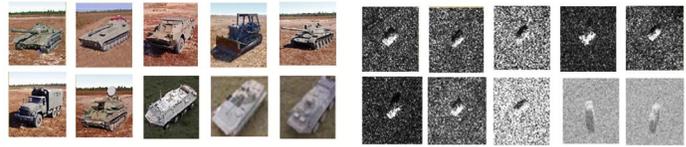

Fig. 5: MSTAR sensitive data used by a CDL victim that might be under GAN threat: 10 optical images of sensitive targets (left) and corresponding SAR images (right).

As for types 3 and 5 attacks, they can be attenuated by method for dealing with 'rootkit', i.e. every time the server intends to load the new gradients, it should first load them to a blank model (pre-train model in this case), and later compare the running result with that of a previous model, if during the training, obvious gradient vanishes or explosion problems are found, then these gradients can be classified as type 3 attack, and should be labeled as 'suspicious'. Besides, the participant who uploads these parameters should be considered to be put on the blacklist for double-checking. In addition, when the model performs significantly worse with the newly uploaded parameters, the type 3 attack might also be detected, and relevant participants should be at least isolated with a lower data transmission rate being assigned (the reason why not putting these participants in the blacklist is that an innocent participant may also upload bad gradients by mistakes. Therefore, this type of attack should be further scrutinized. For type 5 attack, the illegal parameters alternation can be prevented by the firewall layer $W_f$, as once the backpropagation starts, weights of $W_f$ will be changed. As a result, the 100% dropout can work to prevent the adversary from further attacking the inner core layers of the model.

## V. EXPERIMENT PREPARATION AND PERFORMANCE ANALYSIS

### A. MV-CNN Expandable large Network Architecture for Metaverse-oriented CDL

The whole process of the proposed protocol is verified in a CDL-based network training for the metaverse, which is associated with Multiview-Convolutional Neural Network (MV-CNN) that works as the model is trained and used to load trained parameters uploaded by participants. The reason why this kind of structure is chosen is due to the paralleled structure that is proven to have better performance than a single branch structure [36], while it is also capable of being the network core to support Algorithm 1 due to its quick check ability (will be verified in section V, part D). More importantly, the mult-iview structure can be easily fed with multi-modality data collected from various sensors in the process of constructing a digital circumstance. This merit is helpful in speeding up and realizing a metaverse-oriented application. And this structure has been widely applied in various tasks such as detection, classification, 3D-reconstruction, density estimation [37]–[40].

Algorithm 2 and Fig. 4 introduce the data loading and network construction details of the tested MV-CNN model where a specialized data loader formation method is applied to make the data compatible with the relevant MV-CNN structure. In detail, unlike the traditional <single input, single label>tuple, the MV structure separately loads inputs and later appends them to achieve <multi-inputs, one label>structure. In addition, the network merges all these inputs in sequence, and this network structure is expandable.

### B. MSTAR Data Set

As introduced in previous sections, the CDL can be beneficial in creating the metaverse by distributing the training process to different participants so that more diverse data sets can be applied in network training to improve the generalization power of the model which is required to be fully pre-trained for metaverse usage. In this section, the MSTAR data [41] is used to represent a kind of sensitive data set that one Victim (V, as a participant node of CDL) may use (just like the bust measurement data in the metaverse 3D shopping example). The MSTAR contains ten categories of military targets including two kinds of tanks, four classes of armored carriers, bulldozers, trucks, howitzers, and anti-air units. In Fig. 5, details of the aforementioned targets can be found via the optical-SAR pair pictures. For the MSTAR program, since its data were collected by a plane that scans targets using SAR imagery technology from the top, images taken from different aspect angles can therefore be considered as different views which can be later fed into the multi-branch network.

Besides, as mentioned before, this kind of data set is sensitive enough to represent the case in metaverse-oriented CDL when an innocent participant tries to train the model with his/her data with privacy.

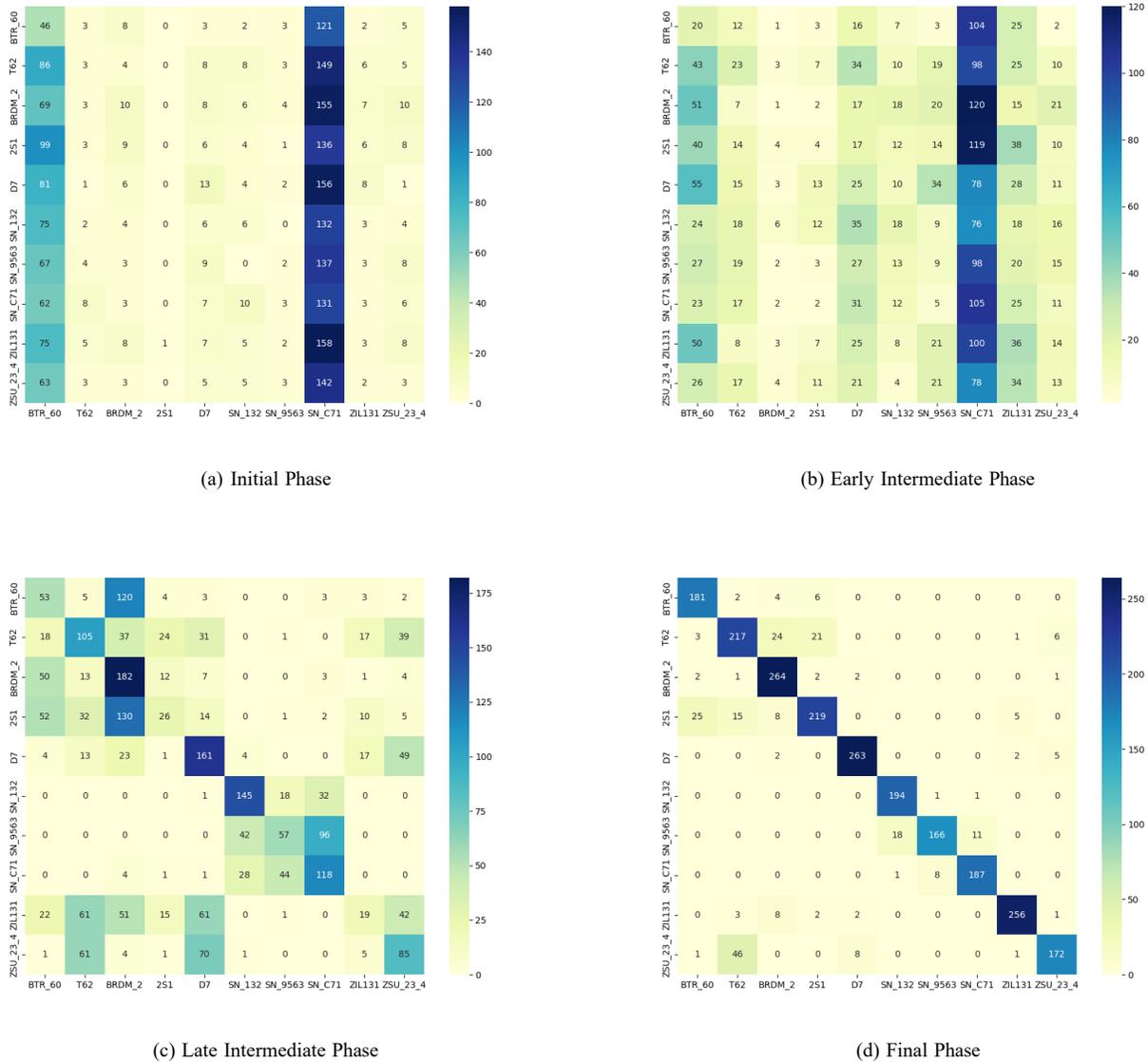

(a) Initial Phase

(b) Early Intermediate Phase

(c) Late Intermediate Phase

(d) Final Phase

Fig. 6: The classification outcomes of various epochs in the model's training process is presented as follows: (a) The initial training status, denoted as epoch=0. (b) The early intermediate status is represented by epoch=40. (c) The late intermediate status, identified by epoch=60. (d) The final stage, is characterized by epoch=75.

### C. Experiment Setup and Classification Performance Analysis

The entire experiment was run on a laptop with AMD Ryzen 9 5900HX, and a 16 GB NVIDIA RTX 3080 Laptop GPU. This experiment has used the 'simple hold-out validation' which splits the train, validation, and test partitions as 7:1:2. Besides, the 'k-fold validation' method is also encouraged. Fig6. shows the classification results of the MV-CNN, where the entire training process is shown as the classification transition of the heatmaps.

Refer to the four selective stages of the whole training phase, it is clear that all inputs are gradually sorted, moving from the initial random status (Fig. 6a), where the majority of inputs were assigned to an arbitrary column, to intermediate stages (Figs. 6b, 6c) where the most predicted results gradually aligned, and ultimately to a well-trained stage, where the majority of predicted results are in diagonal, indicating the final trained model can be capable of accurately recognizing all classes.

The ultimate heatmap in Fig6d illustrates that the majority of predicted labels corresponded accurately to the ground truth, as indicated by the blue or dark blue cells. However, there were a few instances where the model misclassified labels, as evidenced by the lighter cells, also Fig6d, reports a detailed count of each class, enabling further analysis of the model's performance. The achieved highest accuracy during the testing process is 95.58% which serves as a notable accomplishment and a more robust result is 91.83%. It is worth noting that while accuracy is a crucial metric in assessing classification models, other metrics such as precision, recall, and F1-score can provide additional insights into the model's effectiveness which will also be shown in section V, part D.

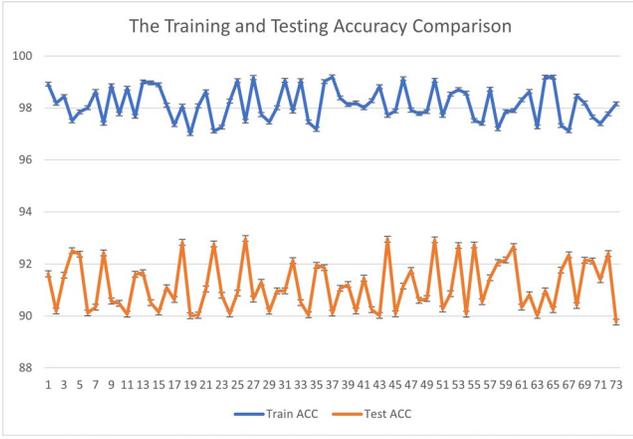

Fig. 7: The record of training and testing accuracy in MV-CNN for CDL after an optimum level has been achieved the first time (represents as epoch 0)

Fig. 7. records the training results after a 'saturation point' of epoch 30, besides, the participant model training process is fully recorded at an epoch level where it can be observed that for the MSTAR data set, a stable training accuracy of 98% is achieved, and median test accuracy of more than 90.85% is obtained.

### D. Model Evaluation and MV-CNN Single-branch Swift Check

The evaluation of models relies on indicators such as Recall (R), Precision (P), and F1-Score (F1) [42]. Specific explanation and mathematical expressions for these parameters can be found in [43], [44]:

$$P_c = \frac{TP_c}{TP_c + FP_c}, P = \frac{\sum_{C=1}^{N} P}{N} \quad (5)$$

$$R_c = \frac{TP_c}{TP_c + FN_c}, R = \frac{\sum_{C=1}^{N} R_c}{N} \quad (6)$$

$$F1_c = \frac{2 \times P_c \times R_c}{P_c + R_c}, F1 = \frac{\sum_{C=1}^{N} F1_c}{N} \quad (7)$$

Where the evaluation of each category is measured by class-separated precision ($P_c$), Recall ($R_c$), and F1-Score ($F1_c$), with each score ranging from 0 to 9. These scores are calculated from the number of false positives (FP), true positives (TP), false negatives (FN), and true negatives (TN). Specifically, FP represents the number of cases where the predicted label is positive while the actual label is negative, TP denotes the number of cases where both the predicted and actual labels are positive, FN represents the number of cases where the predicted label is negative while the actual label is positive, and TN refers to the number of cases where both the predicted and actual labels are negative. And the evaluation

TABLE I: Model Evaluation Details of the Metaverse-oriented MV-CNN from 1-view to 8-view

| Methods | P | R | Acc | F1 | FPS | Params(M) |
|---------|-------|-------|-------|-------|-----|-----------|
| 1-view  | 0.827 | 0.833 | 0.836 | 0.831 | **127** | **6.66** |
| 4-view  | **0.915** | **0.925** | **0.918** | **0.921** | 47  | 19.3 |

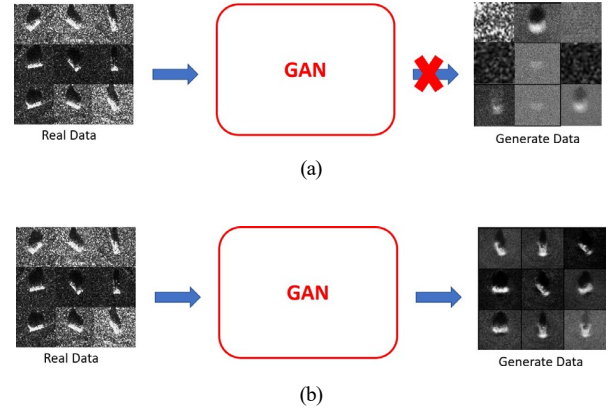

Fig. 8: (a) A successful anti-GAN attack result and (b) a successful GAN attack

results for our model indicate a Recall of 92.5%, a Precision of 91.5%, and an F1-Score of 92.1%.

In Table I, all traditional model evaluation factors of both 1-view and 4-view structures are given, and the Frames Per Second (FPS) information is also given as evidence to show that the 1-view metaverse-oriented CDL model can perform much swifter (2.7 times) than the 4-view together with a lighter weight of 6.66 million (only 35% to that of the 4-view, which is 19.3 million). This fast processing capability can support the fast-check with one-branch feasibility as mentioned in section IV, part A and this one-branch quick check method is even more useful when the network is expanded to 10 or 15 views with millions of parameters.

### E. Successful and Unsuccessful GAN-Attack

This section focuses on illustrating the GAN-attack in the metaverse-oriented CDL launched by an adversary. It is noticeable that since a typical GAN-attack is completed locally with downloaded parameters, the method adopted in this paper to defend GAN-attack is to use Access Credential ($C_{at}$) to limit the request from the adversary which is identified by heuristic scanning. Fig. 8 shows a typical comparison of a successful GAN-attack and an unsuccessful one. From this figure, it is noticed that a successful GAN-attack can generate images that are highly similar to the original ones as the generator always confront (or is taught) by the discriminator and finally outputs a confidence of 0.5, which indicates the discriminator has 50% confidence to say a true image is 'true' (true positive, TP) while another 50% confidence remains for the statement that the true image is 'false' (false negative, FN). This fact indicates that the real image is like a fake image and vice versa. In Fig. 8a, It can be found that the generated data has little difference from the real data and it is not possible for anyone to judge the image if they were mixed together. This fact indicates a successful GAN-attack as the sensitive information of the participant is reproduced. Fig. 8b. shows an unsuccessful GAN-attack where images with full-screen noise are observed in the image either in the bar (the interested target is not reproduced), or in black and grey dots (the GAN gets nothing from the downloaded parameters).

TABLE II: Methods Comparison

| Papers | Pathak et al. [45] | Phong et al. [6] | Yan et al. [16] | Chen et al. [15] | Our work |
|---|---|---|---|---|---|
| Anti-attack strategy | Differential Privacy-based | HE-based | Buried point & lr cut-off | TTP-based | MV-aided aggregation |
| Participant's hostility | innocent | innocent | Semi-hostile | Malicious | Semi-hostile |
| Anti-GAN capability | F | F | T | T | T |
| Scalability | F | T | N/A | N/A | T |
| Privacy-preserving | T | T | T | T | T |

There are three main aspects that could be improved in the future. First, to check the received gradients, this paper activated one branch of the expandable MV-CNN which is proven to be efficient and fast. However, to make the protocol more universally suitable, generalized network cases which has cascade structures such as transformer-based [46]–[48] or YOLO-based structures [49]–[52] should be considered. Besides, since the GAN-attack is a local process and hard to restrict after gradients were downloaded, this paper mainly adopted a heuristic scanning-based method with a credential $C_a t$ limiting the access. However, it is very appreciable to explore some offline methods such as creating special embedded signatures which can destroy gradients once they are used in GAN-attack. Finally, the work can also include a trusted third party so as to protect more safety as well as make the server participate more in the anti-attack process.

### F. Weight Aggregation and Comparison

In CDL server, the weight aggregation is a necessary step to make a model converge, which ensures an iterative update without nodes' direct participation. And using a multiview core is beneficial to enhance privacy protection to this aggregation process. To construct an interactive digital circumstance, a metaverse-oriented CDL needs to be fed with multi-modality data collected from various sensors. Based on this fact, gradients from different sensors will be assigned to different branches, and this separation is helpful to mitigate the risk of information leakage and to reduce the harm that one infected branch can possibly do to the whole model. Moreover, after fast check process, a score will be generated for different participants according to their gradients' performance. And this score will later be altered to a confidence that links to participant's accessibility to this model as shown in Algorithm 2 (Concatenate: $branchN \times confn$).

Table II presents a comparison of our approach to previous works in 5 dimensions. Although all these works provide privacy-preseving services. The major difference is scalability and anti-GAN capability. Previous works such as Pathak [45] and Phong's algorithms [6], although applied effective technologies such as differential privacy or homomorphic encryption-based structures, these technologies mostly focuses on enhancing server's security regarding a malicous node while neglecting the possibility that innocent participants can also be targeted by the attack. Therefore, these methods are limited in defending GAN-based attack especially in an analysis that participants are mostly assumed innocent. On the contrary, in [15], [16] and our method, participants are assumed to be either semi-hostile or active adversaries which will be rejected to have continuous interactions with the server once suspicious behaviours are observed. In terms of scalability, [15], [16], [45] show a CDL-related schematic while did not directly present how a paralleled or distributed structure can be organized either in formulas or algorithms. In [6], the distributed nature of CDL is deduced in formulas, and our method chooes to present this distribution in a structural manner as shown in Algorithm 2 where a MV-aided network core is applied to assign different branches to various modality users. Moreover, this paralleled structure is compatible with the di-centralized essence of CDL, and can be easily expanded to more branches without detriment to the network's capability.

## VI. CONCLUSION

This paper discussed the importance of CDL and its security under the GAN-attack threat. Five types of attacks are listed and a protocol that can process four out of the five threats is given under the assumption of malicious participants being engaged in the local training process. The core methodology was to prevent the reproduction of the inputs of innocent participants by simply downloading the model parameters and using the GAN model. Another security protection aspect was the introduction of the firewall layer in the model to be trained to prevent malicious gradients or even bad parameters from being sent to the model and hence, destroying the training process. The performance of the whole protocol was discussed in steps. Finally, a typical MV-CNN as the CDL model was selected, and the case of locally training it on a special dataset was investigated. Results showed that the entire structure can theoretically perform well in protecting data privacy as well as defending against the GAN-attack by deactivating the illegal access and therefore, making only noise images can be restored by potential adversaries.

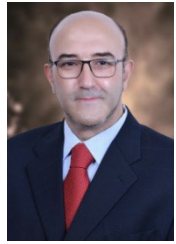
**NAOUFEL WERGHI** (Senior Member, IEEE) received the Ph.D. degree in computer vision from the University of Strasbourg, Strasbourg, France, in 1996. He was a Research Fellow with the Division of Informatics, The University of Edinburgh, Edinburgh, U.K., and a Lecturer with the Department of Computer Sciences, University of Glasgow, Glasgow, U.K. He was a Visiting Professor with the Department of Electrical and Computer Engineering, University of Louisville, Louisville, KY, USA. He is currently a Full Professor with the Department of Electrical and Computer Engineering, Khalifa University, Abu Dhabi, United Arab Emirates. His research interests include image analysis and interpretation, where he has been leading several funded projects in the areas of biometrics, medical imaging, and intelligent systems.

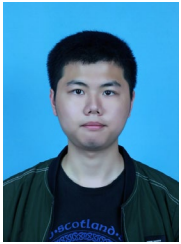
**PENGFEI LI** received his Bachelor of Engineering (B.Eng.) with Honours of the First Class in Electronic and Electrical Engineering from the University of Glasgow (UOG), UK, in 2022, and received his another B.Eng. degree in Electronic and Information Engineering from the University of Electronic Science and Technology of China (UESTC), China. He is currently pursuing a master's degree in electrical and computer engineering with Khalifa University of Science and Technology (KUST), United Arab Emirates. He is also seeking a potential PhD position. His research interests include temporal event detection, multi-scenario segmentation, uncertainty evaluation with deep learning, cryptographic protocol, and explainable artificial intelligence.

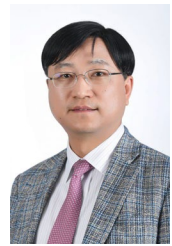
**CHAN YEOB YEUN** (Senior Member, IEEE) received the M.Sc. and Ph.D. degrees in information security from Royal Holloway, University of London, in 1996 and 2000, respectively. After his Ph.D. degree, he joined Toshiba TRL, Bristol, U.K., and later, he became the Vice President at the Mobile Handset Research and Development Center, LG Electronics, Seoul, South Korea, in 2005. He was responsible for developing mobile TV technologies and related security. He left LG Electronics, in 2007, and joined ICU (merged with KAIST), South Korea, until August 2008, and then the Khalifa University of Science and Technology, in September 2008. He is currently a Researcher in cybersecurity, including the IoT/USN security, cyber-physical system security, cloud/fog security, and cryptographic techniques, an Associate Professor with the Department of Electrical Engineering and Computer Science, and the Cybersecurity Leader of the Center for Cyber-Physical Systems (C2PS). He also enjoys lecturing for M.Sc. degree in cyber security and Ph.D. degree in engineering courses at Khalifa University. He has published more than 140 journal articles and conference papers, nine book chapters, and ten international patent applications. He also works on the editorial board of multiple international journals and on the steering committee of international conferences.

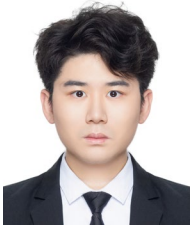
**ZHIBO ZHANG** received the Bachelor of Science degree in mechatronics engineering from Northwestern Polytechnical University, China, in 2021. He is currently pursuing a master's degree in electrical and computer engineering at Khalifa University, United Arab Emirates. He was awarded the Award for Outstanding Young Researcher by Khalifa University in 2022. His research interests focus on biometrics, cyber-physical system, cyber security, Explainable Artificial Intelligence, and Federated Learning.

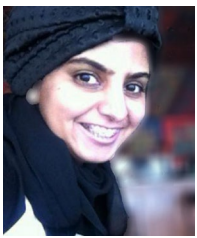
**AMEENA SAAD AL-SUMAITI** (Senior Member, IEEE) received the B.Sc. degree in electrical engineering from United Arab Emirates University, UAE, in 2008, and the M.A.Sc. and Ph.D. degrees in electrical and computer engineering from the University of Waterloo, Canada, in 2010 and 2015, respectively. She was a Visiting Assistant Professor with MIT, Cambridge, MA, USA, in 2017. She is currently an Associate Professor with the Department of Electrical and Computer Engineering, Khalifa University, Abu Dhabi, UAE. Her research interest includes intelligent systems, energy economics, energy policy, and cyber-physical systems.